\documentstyle[12pt,aaspp4]{article}

\begin{document}
\title{Isolated versus Common Envelope Dynamos in Planetary Nebula Progenitors}
\author{J. Nordhaus\altaffilmark{1,2}, E. G. Blackman\altaffilmark{1,2}, A. Frank\altaffilmark{1,2}}
\affil{1. Dept. of Physics and Astronomy, Univ. of Rochester,
    Rochester, NY 14627}\affil{2. Laboratory for Laser Energetics, Univ. of Rochester, Rochester, NY 14623}
\begin{abstract}
The origin, evolution and role of magnetic fields in the production and shaping of proto-planetary  and planetary nebulae (PPNe, PNe) is a subject of active research.  Most PNe and PPNe are axisymmetric with many exhibiting highly collimated outflows, however, it is important to understand whether such structures can be generated by isolated  stars or require the presence of a binary companion.  Toward this end we study a dynamical, large-scale $\alpha-\Omega$ interface dynamo operating in a $3.0$ $M_\odot$ Asymptotic Giant Branch star (AGB) in both an isolated setting and one in which a low-mass companion is embedded inside the envelope.  The back reaction of the fields on the shear is included and differential rotation and rotation deplete via turbulent dissipation and Poynting flux.  For the isolated star, the shear must be resupplied in order to sufficiently sustain the dynamo.  Furthermore, we investigate the energy requirements that convection must satisfy to accomplish this by analogy to the sun.  For the common envelope case, a robust dynamo results, unbinding the envelope under a range of conditions.  Two qualitatively different types of explosion may arise: (i) magnetically induced, possibly resulting in collimated bipolar outflows and (ii) thermally induced from turbulent dissipation, possibly resulting in quasi-spherical outflows.  A range of models is presented for a variety of companion masses.  

\end{abstract}
\section{Introduction}

Most planetary and proto-planetary nebulae (PNe, PPNe) are highly aspherical and exhibit a diversified morphology of axisymmetric structures and/or collimated jets.  However, the production of such richly varied systems and shaping mechanisms, remains an open topic (\cite{2002ARA&A..40..439B}).  A central question is whether a binary is required to produce such asymmetries or if an isolated Asymptotic Giant Branch star (AGB) is sufficient.  

Recently, detection of magnetic fields in AGB stars (\cite{2004MNRAS.348...34E}, \cite{2004MNRAS.354..529B}) and the central stars of PNe (\cite{2005A&A...432..273J}) has sustained interest in magnetic launching and collimation mechanisms.  Observational evidence of a magnetically collimated jet in an evolved AGB star (\cite{2006Natur.440...58V}) further suggests that the magnetic field may play a dynamical role.

Single star magnetic outflow models have been proposed as mechanisms for producing and shaping PPNe and PNe (\cite{1997ApJ...489..946P}, \cite{EB2001}).  Whether such models can power and shape the PPNe is uncertain (\cite{2006PASP..118..260S}).  In particular, envelope dynamos are expected to be short ($<100$ yrs) and drain differential rotation energy rapidly (\cite{2004ASPC..313..401B}), making it unlikely that isolated stars can produce observed asymmetries.  If convection can resupply differential rotation energy, then an envelope dynamo in an isolated AGB star may be viable.  Anisotropic convection in the sun resupplies differential rotation through the $\lambda$-effect (\cite{1989QB523.R93......}, \cite{2004muga.book.....R}), however it is not clear if such a mechanism operates in AGB stars.

Rather than single star models, the observed asymmetries may instead be the result of energy and angular momentum supplied by binary interactions.  This is supported by recent surveys suggesting that most, if not all PNe involve binary systems (\cite{2004ApJ...602L..93D}, \cite{2004ASPC..313..515S}, \cite{2006astro.ph..6354M}, \cite{2006A&A...452..257M}). Although there are many different types of binary interactions and outcomes, here we focus on common envelope (CE) evolution in the context of PNe progenitors (\cite{1993PASP..105.1373I}).  A common envelope (CE) model in which low-mass ($<0.3$ $M_\odot$) companions were embedded into the envelope of a $3.0$ $M_\odot$ AGB star was investigated in \cite{JN2006}.  The CE evolution is advantageous as it can supply angular momentum in an extremely short period ($<1$ yr) and produce a range of PPNe outflows.  Such a model would predict white dwarf + brown dwarf close binaries that survive the CE phase.  Recently, a WD+ BD binary in a close orbit has been detected (\cite{2006Natur.442..543M}, \cite{2006astro.ph..9366B}).  A separation distance of $0.65$ $R_\odot$ indicates that the system incurred a common envelope phase in which the brown dwarf was responsible for ejecting the progenitor envelope.  This further motivates more detailed study of CE induced PNe.

In this paper, we reinvestigate the magnetic model presented in \cite{EB2001} in more detail, in an effort to determine the viability of a single star dynamo.  We compare the results to a model in which the rotation profile is supplied by a CE phase as in \cite{JN2006}.  In section 2, we review previous work, compare single star evolution to that resulting from a CE interaction and focus on the depth that the poloidal field can diffuse into the shear zone.  We present our interface dynamo in section 3, including the detailed back reaction of the fields on the shear, the generation of heat through turbulent dissipation and the spin down of the star due to Poynting flux.  In Section 4, we present the results of our model for three cases:  (i)  an isolated dynamo in which convection does not resupply differential rotation, (ii) an isolated dynamo in which a fraction of the convective energy resupplies differential rotation energy and (iii) a dynamo resulting from the in-spiral of a low-mass ($\leq0.05$ $M_\odot$) companion inside the stellar envelope.  We conclude in section 5.

\section{Dynamos, Common Envelopes and Isolated AGB Evolution}

A central issue in magnetic PNe progenitor models is whether an isolated AGB star can sustain the necessary field strengths and corresponding Poynting flux to unbind the stellar envelope and produce collimated outflows.  Several authors have appealed to various mechanisms with which to produce magnetically mediated outflows in isolated settings (\cite{1992MNRAS.256..269T}, \cite{1993JApA...14...65P}, \cite{1997ApJ...489..946P}, \cite{EB2001}, \cite{2002MNRAS.329..204S}, \cite{2004ASPC..313..449M}).  \cite{2002MNRAS.329..204S} appeal to an $\alpha^2-\Omega$ dynamo operating in the AGB envelope as a means of enhancing dust formation near magnetic cool spots on the stellar surface.  The corresponding fields are not strong enough to dynamically alter the geometry, however enhanced mass-loss near the spots could form an elliptical PN.  Such a model does not produce strongly bipolar PNe. 

Other authors have investigated $\alpha-\Omega$ dynamo models which tap into the differential rotation energy reservoir between core and convective zone.  \cite{1997ApJ...489..946P} solved for a steady-state, radial dynamo model from inside the AGB core to produce fields throughout the envelope.   Toroidal field strengths of $10^6$ G are obtained at the surface of the core with poloidal field strengths about an order of magnitude lower.  No back-reaction of the fields on the rotation profile was included.

\cite{EB2001} investigated a simplistic interface dynamo model (\cite{1993ApJ...408..707P}, \cite{1995ApJ...453..403T}) operating at the base of the convective zone in our $3$ $M_\odot$ AGB star (see Fig. \ref{geometry}).  Angular momentum is conserved on spherical shells as the star evolves off the main sequence and the resulting rotation profile is used to calculate field strengths.  To drive PPNe, the corresponding dynamo must operate through the entire lifetime of the AGB phase ($\sim10^5$ yrs) until radiation pressure has bled most of the envelope material away.  Only then can the Poynting flux unbind the remaining material.  But, there are challenges for this model.  The differential rotation zone is chosen to be $\sim1/2$ of the total distance from core to convective zone and $\sim1/2$ the length of the convective region.  This results in only a small fraction of the free shear energy available for field amplification.  The majority of the shear energy is located deeper, near the core-envelope boundary.  The actual thickness of the differential rotation layer tapped by the dynamo is uncertain.  If more of the shear energy were extracted, the envelope could be blown off too early.  On the other hand, any dynamo operating beneath the envelope would drain differential rotation on time scales short compared to the AGB lifetime.  Only if the differential rotation is re-seeded might this problem be overcome.  Even in the sun, the complex interaction between anisotropic convection and the resupply of differential rotation is not fully understood.  Recently, the role of downward pumping and penetration depth in the solar tachocline has been investigated (\cite{2006ApJ...648L.157B}, \cite{2006ApJ...638..564D}).


If isolated star models fail to generate sufficient Poynting flux, common envelope evolution provides an alternative mechanism with which to supply significant differential rotation energy over very short periods ($\sim1-10$ yrs). The in-spiral of a low-mass secondary ($<0.3$ $M_\odot$) through the envelope of a $3.0$ $M_\odot$ star in the AGB phase was investigated in \cite{JN2006}.  Three qualitative scenarios were found dependent on the mass of the companion: (i) direction ejection of envelope material resulting in an equatorial outflow, (ii) spin-up of the envelope resulting in an explosive dynamo driven jet along the rotational axis and (iii) tidal shredding into a disc which facilitates a jet.  In this paper, we investigate (ii) further, in addition to presenting results for an isolated star dynamo.

\section{Dynamical Equations}
In order to determine the temporal behavior of the large-scale magnetic field, we employ the mean-field induction equation which results from averaging the standard induction equation in the presence of helical velocity fluctuations.  The result is (\cite{1979cmft.book.....P})

\begin{eqnarray}
\partial_t\bf{\overline{B}}&=&\nabla\times{\bf \mathcal E}
+\nabla\times\left(\bf{\overline{U}}\times\bf{\overline{B}}\right)+\lambda{\nabla^{2}}\bf{\overline{B}} \\
& = &\nabla\times(\alpha\bf{\overline{B}})+\nabla\times\left(\bf{\overline{U}}\times\bf{\overline{B}}\right)+\nabla\times\left(\beta\nabla\times\bf{\overline{B}}\right)+\lambda{\nabla^{\rm 2}}\bf{\overline{B}}\nonumber
\end{eqnarray}
where $\overline{\bf{U}}$ is the mean velocity field, $\overline{\bf{B}}$ the mean magnetic field, $\lambda$ the micro-physical magnetic diffusivity, $\beta$ the turbulent diffusion such that $\lambda\ll\beta$, $\alpha$ the pseudo-scalar helicity coefficient, and $\mathcal{E}=<\bf{u}\times\bf{b}>=\alpha\bf{\overline{B}}+\beta\nabla\times\bf{\overline{B}}$ the turbulent electromotive force (\cite{1978mfge.book.....M}).

Although we envision the dynamo engine operating in a spherical or quasi-spherical AGB star, for present purposes we work in local Cartesian coordinates.  Such interface dynamo models have been employed in a variety of systems ranging from late type stars (\cite{1982A&A...108..322R}) to white dwarfs (\cite{1994ApJ...430..834M}) to supernova progenitors (\cite{BNT}).  The coordinate system and global geometry is presented in Fig. \ref{geometry}.  The convection zone extends from the stellar surface to the interface at $r=r_c$.  In this layer, convective twisting motions convert buoyant toroidal field into poloidal field through the $\alpha$-effect.  Below the convection layer, the differential rotation zone shears poloidal field back into toroidal field through the $\Omega$-effect.  Defining the vector potential as ${\bf\overline{A}}=(\overline{A}_x,\overline{A},\overline{A}_z)$ and decomposing the mean field into toroidal and poloidal components, ${\bf\overline{B}}=(0,\overline{B},\partial_x\overline{A})$, generates two coupled equations for the time evolution of both components of the magnetic field.  

In order to capture aspects of the 2-D geometry within the framework of our simple Cartesian 0.5-D model, we break the turbulent diffusion into two distinct values: $\beta_p$, corresponding to the poloidal field which grows primarily in the convective region and $\beta_\phi$, corresponding to the toroidal field which is amplified in the differential rotation zone.  We also employ $\beta_\phi$ as the turbulent diffusion coefficient for the toroidal velocity.  Since the convective region is highly turbulent and the differential rotation zone is more weakly turbulent, we have that $\beta_\phi\ll\beta_p$.  The ratio of these can be defined as the turbulent magnetic Prandtl number as $Pr_{p}\equiv\frac{\beta_\phi}{\beta_p}$.  Then, assuming axisymmetry (i.e. $\partial_y\bf{\overline{S}}=0$) for all mean quantities $\bf{\overline{S}}$ and defining the velocity field as ${\bf\overline{U}}=(0,\overline{U},u)$, we obtain

\begin{equation}
\partial_t\overline{B}=-\alpha\partial_x^2\overline{A}-\partial_z\alpha\partial_x\overline{A}+\partial_x\overline{A}\partial_z\overline{U}-u\partial_z\overline{B}-u\overline{B}/{L}+\beta_\phi\left(\partial_x^2+\partial_z^2\right)\overline{B}
\end{equation}
\begin{equation}
\partial_t\overline{A}=\alpha\overline{B}+\beta_p\left(\partial_x^2+\partial_z^2\right)\overline{A}.
\end{equation}
where $\overline{B}\partial_zu\sim{u}\overline{B}/L$ represents a buoyant loss of magnetic flux and $u>0$ (to be made explicit later).  We further assume that the Fourier transforms of the fields are proportional to delta functions implying that the mean-field has one large scale.  We correspondingly define $\left[\overline{B},\overline{A}\right]=\left[B(t),A(t)\right]e^{i\left(k_xx+k_zz\right)}$ where $A(t)$ and $B(t)$ are complex valued functions of time.  Then, setting $k^2=k_x^2+k_z^2$ and using $\partial_z\overline{U}\simeq-r_c\Delta\Omega/L$ yields the following

\begin{equation}
\partial_tB=\alpha{k_x}^2A-ik_xr_cA\frac{\Delta\Omega}{L}-iuk_zB-uB/L-\beta_\phi{k^2}B
\label{toroidal}
\end{equation}
\begin{equation}
\partial_tA=\alpha{B}-\beta_pk^2A,
\label{polidal}
\end{equation}
where the rotation profile across the differential rotation layer varies from $\Omega$ at the interface to $\Omega+\Delta\Omega$ at $r_c-L$.  Thus, $\Delta\Omega$ is a measure of the shear in the differential rotation zone.  If $\Delta\Omega=0$, then the system exhibits solid body rotation.  In addition, we parameterize  the turbulent diffusion coefficients as $[\beta_\phi,\beta_p]=[c_\phi,c_p]{vL_1}$, where $c_\phi$ and $c_p$ are distinct dimensionless constants and $v$ is a typical convective velocity in the $\alpha$-layer.

For the loss of toroidal flux due to magnetic buoyancy, we use the following expression for the rise velocity of a magnetic flux tube (\cite{1955ApJ...121..491P}, \cite{1995ApJ...453..403T}):

\begin{equation}
u=\frac{3Q}{8}\left(\frac{a}{L}\right)^2\frac{|\overline{B}|^2}{4\pi\rho{v}}=\frac{3Q}{32}\frac{\overline{V_A}^2}{v}
\end{equation}
where $a$ (assumed to be $L/2$) is the radius of the flux tube, $\overline{V}_A$ is the Alfv\'{e}n velocity associated with the large scale field and $Q$ is a dimensionless constant of order unity.

\subsection{Evolution of $\Omega$ and $\Delta\Omega$}
As differential rotation energy is tapped by field amplification, the corresponding Poynting flux drains rotational energy at the interface.  In addition, turbulent diffusion converts differential rotation energy into heat which may also be used to unbind the AGB envelope.  Therefore, to investigate the interaction between field amplification, differential rotation and rotation, we derive equations for the evolution of $\Delta\Omega$ and $\Omega$.  The mean-field Navier-Stokes equation is given by

\begin{equation}
\partial_t\bf{\overline{U}}=-\bf{\overline{U}}\cdot\nabla\overline{U}+\frac{1}{4\pi\rho}\left(\bf{\overline{B}}\cdot\nabla\right)\overline{B}+\beta_\phi\nabla^2\overline{U}
\end{equation}
where $\rho$ is the fluid density, $\beta_\phi$ the turbulent viscosity and we have assumed that $b\ll\overline{B}$ and $u\ll\overline{U}$ in the weakly turbulent shear layer.  Then, taking the $\hat{y}$-component (see Fig. \ref{geometry}) yields the following

\begin{equation}
\partial_t\overline{U}\simeq\frac{1}{4\pi\rho}\partial_x\overline{A}\partial_z\overline{B}+\beta_\phi\left(\partial_x^2+\partial_z^2\right)\overline{U}.
\end{equation}
Using the fact that, $\partial_z\overline{U}\simeq-r_c\Delta\Omega/L$, we can link $\Delta\Omega$ with $\overline{U}$ as follows
\begin{equation}
-\partial_t\left(r_c\Delta\Omega\right)=\partial_t\left(\overline{U}\left(r_c\right)-\overline{U}\left(r_c-L\right)\right)\simeq\partial_t\left(L\partial_x\overline{U}\right).
\end{equation}
Subtracting the time-dependent velocity equation at $r_c-L$ from $r_c$ and using the relation $\partial_t\overline{U}|_{r_c}-\partial_t\overline{U}|_{r_c-L}\sim{L}\partial_z\partial_t\overline{U}|_{r_c}$ yields the following

\begin{equation}
\partial_t\Delta\Omega=\frac{L}{4\pi{r_c}\rho}\left[-k_xk_z^2\left(Re(\overline{A})Re(i\overline{B})+Re(i\overline{A})Re(\overline{B})\right)-\frac{\partial_z\rho}{\rho}{k_x}k_zRe(i\overline{A})Re(i\overline{B})\right]-\frac{\beta_\phi}{L^2}\Delta\Omega,
\label{dr}
\end{equation}
where we have assumed $\rho=\overline{\rho}$ and thus $\partial_z\rho\sim\left(\rho_2-\rho_1\right)/L$ is the change in density across the shear layer.

In addition to the dynamic shear, we allow for the rotational energy of the field-anchored matter to drain via Poynting flux.  No appreciable toroidal field amplification occurs above the interface, so we calculate the Poynting flux at the base of the convection zone.  The total integrated Poynting flux at $r_c$ is given by
 
\begin{eqnarray}
L_{mag}&=&\frac{c}{4\pi}\int\left(\bf\overline{E}\times\bf\overline{B}\right)\cdot{\bf{dS_c}} \nonumber\\
&\simeq&-Re(\overline{B})Re(\partial_x\overline{A})\Omega{r_c^3}.
\end{eqnarray}
When the toroidal and poloidal fields are out of phase, there is a maximum magnetic luminosity that is not the respective product of the maximum individual field strengths (\cite{BNT}).  This is a generic feature of our interface dynamo solutions and will be presented in Section 4.

To arrive at a dynamical equation for $\Omega$, we must calculate the available rotational energy in the shear layer.  We estimate the rotational energy in the differential rotation zone as $E_{rot}\sim{M}_{\Delta\Omega}r_c^2\Omega^2/2$.  Various mass secondaries can supply a range of differential rotation in addition to rotational energy.  For a $0.05$ $M_\odot$ brown dwarf, the total shear energy from the convective boundary to the tidal shredding radius is $\sim5\times10^{47}$ ergs in the AGB envelope.  This is approximately 4 times the binding energy of the entire AGB envelope.  As the star evolves  into its thermal pulsing phase, the convective zone deepens and both the thickness and mass of the shear layer shrink.  It may also be possible to power an interface dynamo during a later phase in the stars evolution.  In this paper, we focus on the beginning of AGB phase.

Equating the time derivative of the rotational energy with the magnetic luminosity gives

\begin{equation}
\partial_t\Omega\simeq\frac{Re(\overline{B})Re(\partial_z\overline{A})Lr_c}{M_{\Delta\Omega}\delta}.
\label{omega}
\end{equation}
In arriving at Eq. (\ref{omega}), we have multiplied the available rotational energy by a factor of $\delta/L$.  The penetration length, $\delta$, represents the depth at which the poloidal field can diffuse into the shear layer.  If $\delta/L=1$, the total shear energy in the differential rotation layer is available for extraction.  We estimate the penetration depth as the distance that the poloidal field can diffuse into the shear layer during a cycle period.  In the kinematic limit, the cycle period is given as $\tau=2\pi\left(\frac{2L}{\alpha_0\Delta\Omega_0kr_c}\right)^{\frac{1}{2}}$ where $\alpha_0$ and $\Delta\Omega_0$ are the initial values (\cite{BNT}).  The cycle period does increase in the dynamical regime, however it does not change appreciably from its initial value.  Therefore, $\tau$ serves as a lower limit for the cycle period.  Correspondingly, we define the penetration depth as 

\begin{equation}
\delta\simeq(\beta_\phi\tau)^{\frac{1}{2}}.
\end{equation}

\subsection{Evolution of $\alpha$}
In addition to a dynamical equation for shear, $\alpha$ quenching can be understood through magnetic helicity conservation (\cite{Kleeorin1982}, \cite{2002PhRvL..89z5007B}, \cite{2005PhR...417....1B}).  In the absence of boundary terms, magnetic helicity is well conserved and the build up of ${\bf\overline{A}\cdot\overline{B}}$ corresponds to a build up of large-scale field.  The small-scale helicity then grows to equal and opposite magnitude of the large-scale helicity.  To maintain simplicity, here we appeal to a parametrized form of $\alpha$ which approximates the non-linear quenching (\cite{BNT}).  We adopt the following profile

\begin{equation}
\alpha=\alpha_0\left(\Omega/\Omega_0\right)Exp\left[-\gamma\frac{\overline{B}^2/8\pi}{\rho_1v^2/2}\right]
\end{equation}
 where $\alpha_0\equiv{c}_\alpha\frac{L_1^2\Omega_0}{r_c}$, $c_\alpha$ a dimensionless constant, $\Omega_o$ the initial rotation rate at the interface and $\rho_1$ the density in the middle of the convective region.

\section{Numerical Results}
To investigate various interface dynamo configurations, Eqs. (4), (5), (10) and (12) are solved numerically, the solutions of which represent the time evolution of $B_\phi$, $B_p$, $\Delta\Omega$ and $\Omega$.  In each case, we employ a 1 G seed field for the real components of both the toroidal and poloidal field.  The fields grow until they are quenched through a drain of the available differential rotation energy.  In all cases, the saturated dynamo is $\Omega$-quenching dominated and not $\alpha$-quenching limited in contrast to \cite{EB2001}.  

We focus on three types of shell dynamos: (i) that of an isolated AGB star, (ii) that of an isolated AGB star in which convection resupplies differential rotation, (iii) that of an AGB star which has been spun up by a companion in-spiraling through the stellar envelope.  We use independent radial rotation profiles for the above cases to calculate the initial value of $\Delta\Omega$.  For the isolated AGB star, we assume that as the star evolves off the main sequence, angular momentum is conserved on spherical shells yielding a rotation profile $\propto{r^{-2}}$.  We consider this case in detail in Section 4.1.  Because convective energy may resupply differential rotation energy analogous to what occurs in the sun, we consider a range of resupply rates in Section 4.2.  In Section 4.3, we consider a rotation profile generated from the in-spiral of a low-mass secondary through the stellar envelope.  The in-spiral time is short compared to the AGB lifetime and rapidly creates a strong shear region beneath the convective zone.  In addition, the in-spiral time is less than or equal to a cycle period, $\tau$, so the angular momentum and energy are supplied to the dynamo on time scales short compared to its growth time.

\subsection{Isolated Dynamo Without Reseeding $\Delta\Omega$}
It has been suggested (\cite{EB2001}), that isolated AGB stars may be able, through dynamo action, to unbind their envelopes and produce collimated outflows.  As the star expands onto and through the AGB branch, the core contracts while the envelope expands, creating a shearing profile throughout the  interior.  A differential rotation zone, coupled with the above convective region provides conditions for large-scale field amplification.  However, field growth requires draining differential rotation energy.  In the sun, the rotation profile is re-established through a transfer of convective energy by way of the $\lambda$-effect (\cite{2004muga.book.....R}).  This results in a steady-state rotation profile and a magnetic cycle with quasi-steady peak field.  However, if convection does not resupply differential rotation energy, any resultant dynamo would be a transient phenomena.

\cite{EB2001} investigated an isolated shell dynamo operating in the AGB phase of our model star.  The depth of the differential rotation zone was taken to be $\sim1/2$ the distance from the base of the convective zone to the stellar core.  In addition, $\Delta\Omega$ and $\Omega$ were assumed to be constant and independent of the magnetic field.  Thus, the dynamo lasted indefinitely, and sustained a toroidal field of $\sim5\times10^4$ G at the interface.  The arbitrarily long lifetime was essential because in order for the large-scale field to drive a self-collimated outflow, the dynamo must last until the end of the AGB phase after the star has radiatively bled most of its envelope material.  To alleviate the assumption of steady $\Delta\Omega$ and $\Omega$ and to study the backreaction of field growth on the shear, we apply our dynamical dynamo model using parameters in \cite{EB2001}. 

We fix the thickness of the differential rotation zone ($L=4.6\times10^{10}$ cm), the rotational speed at the interface ($\Omega_0=5\times10^{-6}$ rad s$^{-1}$), and the shear profile across the $\Omega$-layer.  We then lower the value of $\beta_\phi$.  This lengthens the dynamo lifetime and determines the relative fraction of energy deposited into magnetic or heat sinks.  Fig. \ref{Nature2} shows the isolated interface dynamo for two different values of $\beta_\phi$.  In both cases the depth of the differential rotation zone is fixed and the shear energy in that zone is available for field amplification.  The toroidal and poloidal fields are out of phase and thus generate an oscillatory Poynting flux.  Maximum toroidal field strengths of $\sim5\times10^{3}$ G are obtained, while the poloidal field is significantly lower, with values of $B_p\sim2\times10^2$ G at the interface.  In both cases, the time integrated Poynting flux and turbulent dissipation generate $\sim10^{-5}$ times the binding energy of the envelope.  Therefore, the current parameters can not produce a dynamically influential magnetic outflow or significant heating from turbulent dissipation in the shear zone.  If an isolated interface dynamo were to be viable, a deeper shear zone would be required.  

A higher $c_\phi$ allows the poloidal component of the field to diffuse deeper into the toroidal zone, thus extracting more differential rotation energy.  Hence, the depth of the shear layer is determined by the value of $c_\phi$.  If we fix $L$ as the distance from the base of the convective zone to the core, then the corresponding fields diffuse all the way to the stellar core for $Pr_p$ ${\geq}$ $10^{-6}$.  This unbinds the envelope at the beginning of the AGB phase.  In addition, we can constrain how far the fields would have to diffuse in order to unbind the AGB envelope.  If the poloidal field diffuses to $\sim5\times10^9$ cm within the stellar interior, then $M_{\Delta\Omega}=1.7\times10^{31}$ g and $\Delta\Omega=1.5\times10^{-3}$ rad s$^{-1}$.  This corresponds to $E_{\Delta\Omega}\sim1.5\times10^{47}$ ergs which is comparable to the binding energy of the envelope.  However, in these cases, the dynamo would blow off the envelope prematurely and such a circumstance would contradict observations of AGB lifetimes $\simeq10^5$ yrs.  If an isolated shell dynamo is to be consistent with observations, the dynamo must be  sustained during the end of the AGB phase.

\subsection{Isolated Dynamo With Reseeding $\Delta\Omega$}
Convection might reinforce differential rotation analogous to what occurs in the sun.  Thermal energy and a negative entropy gradient drive convective turbulence.  In the Kolmogorov framework, energy from the large-scale to the dissipative scale cascades at a rate given by
\begin{equation}
\frac{\partial\epsilon}{\partial{t}}\sim\frac{v_l^3}{l}=D
\end{equation}
where $l$ is a length scale, $v_l$ the corresponding velocity and $D$ is independent of scale (\cite{1992phas.book.....S}).  For our PNe progenitor, we use the large-scale convective velocity, $v=v_l=10^5$, cm/s corresponding to the approximate thickness of the convective zone, $l=L_1$.  We envision adding a resupply term to Eq. (10), in which a fraction, $f$, of the turbulent energy cascade rate resupplies shear.  To determine this term, we equate (15) with the time rate of change of kinetic energy in the convective zone and arrive at
\begin{equation}
\frac{\partial\Delta\Omega_{c}}{\partial{t}}=f\left(\frac{M_c}{M_{\Delta\Omega}}\right)\left(\frac{v^3}{L_1L^2\Delta\Omega}\right),
\end{equation}
where $\Delta\Omega_{c}$ is the shear that would be resupplied by the turbulent cascade.  By adding the right side of Eq. (16) to Eq. (10), we can investigate the full range of convective resupply scenarios from $f=0$ (no convective resupply) to $f=1$ (maximum convective resupply).  In addition, we fix the following values:  $\Delta\Omega_0=1.5\times10^{-5}$ rad/s, $\Omega_0=5\times10^{-6}$ rad/s, $L=4.6\times10^{10}$ cm, $c_p=10^{-2}$ and $Q=5.0$.

We consider two sub-scenarios for the convective resupply dynamo: $\Omega$ dynamically evolving and $\Omega$ constant.  In the first case, Poynting flux is allowed to spin down the envelope at the interface.  This requires magnetic buoyancy, which appears in Eq. (4) as terms proportional to $u\overline{B}/L$.  The left graph in Fig. \ref{reseed1} presents the result for dynamical $\Omega$.  Rotation is drained by Poynting flux and we take the most extreme best case of 100 \% ($f=1$) of the turbulent energy cascade rate resupplying differential rotation energy.  This does establish a constant $\Delta\Omega$.  However, the bulk of the Poynting flux is drained in a short burst ($\leq10$ yrs).  It is unfeasible to generate the requisite energy required to produce a magnetically dominated explosion in this case.

For the second case ($\Omega$ constant), we consider the possibility that the field is stored in the interface layer until the corresponding aggregate Poynting flux is able to blow up the envelope through a magnetic "spring" effect.  If the field is trapped, Poynting flux does not emerge from the layer and thus does not spin down the envelope.  Even though the dynamo equations for the two cases are mathematically identical, in the case of steady rotation terms proportional to $u\overline{B}/L$ can be interpreted as diffusion rather than buoyant loss.  The right plot in Fig. \ref{reseed1} shows a constant $\Omega$ with $f=0$ (no convective resupply).  Constant rotation results in a decay of the differential rotation energy.  However, since $\Omega$ is constant, the $\alpha$-effect is non-zero and thus is continually pumping poloidal field into the shear layer.  The achieved poloidal field strength is negligible and energetically insignificant.  It cannot blow off the envelope.

Finally we consider the case in which $\Delta\Omega$ is resupplied and $\Omega$ is constant.  Fig. \ref{reseed2} demonstrates that convective resupply coupled with a constant $\Omega$ results in a sustained Poynting flux.  The Poynting flux of $\sim5\times10^{34}$ erg/s sustained over a period of $10^5$ yrs is enough to overcome the binding energy of the envelope and produce a magnetically driven envelope expulsion.  For $c_\phi=10^{-5}$, $0.1\%$ of the energy cascade rate has been used to resupply the differential rotation and marks the approximate minimum threshold fraction required to blow off the envelope. If $c_\phi\geq5\times10^{-5}$, then the rate of heat produced from turbulent dissipation is greater than the Poynting flux, resulting in thermally induced envelope expulsion.  

Although it is not known whether a mechanism for resupplying differential rotation operates in AGB stars (e.g. \cite{2004muga.book.....R}), we have demonstrated that such an effect may facilitate a dynamo driven envelope expulsion, provided both $\Omega$ and $\Delta\Omega$ are sustained.

\subsection{Common Envelope Dynamo}
In a close binary system, Roche lobe overflow can result in both companions immersed inside a common envelope (\cite{1976IAUS...73...75P}, \cite{1993PASP..105.1373I}).  A drag force due to the velocity difference between companion and envelope, induces in-spiral of the secondary.  As a result, orbital energy and angular momentum are transfered from companion to common envelope.  For low-mass secondaries, the in-spiral time can be extremely fast ($\leq$ 1 yr) supplying the requisite angular momentum on time scales much shorter then the AGB lifetime (\cite{JN2006}).  As the companion traverses the envelope, the transfer of orbital energy alone may be enough to unbind it.  However, if the companion cannot supply the necessary orbital energy, in-spiral continues until the secondary is tidally shredded into a disk.  In addition, angular momentum transfer spins up the envelope resulting in a differentially rotating stellar interior.  Assuming that the AGB star is initially stationary, we can calculate the corresponding rotation profile generated by in-spiral of a companion.  The change in orbital energy, in virial equilibrium, of the secondary is given as
\begin{equation}
\Delta{E_{orb}}(r)=\frac{GM_Tm_2}{2r_\star}-\frac{GMm_2}{2r},
\end{equation}
where $M_T$ is the total mass of the star, $m_2$ the secondary mass, $M$ the enclosed mass at position r, and $r_\star$ is the stellar radius.  A fraction, $\alpha\Delta{E_{orb}}$ of the orbital energy released by the companion is available for mass ejection.  If $\alpha\Delta{E_{orb}}\geq{E_{bind}}$, where $E_{bind}$ is the envelope binding energy, then the secondary has expelled the envelope.
  In order to calculate the rotation profile of the AGB star, we use the gravitational potential energy released by the secondary to spin-up spherical shells as follows:
\begin{equation}
\alpha{\Delta{E_{orb}}}=I_s\Omega_s^2\left(r_i\right),
\end{equation}
where $I_s$ is the moment of inertia of a thin shell, $\Omega_s$ the angular velocity of the shell and $r_i$ the outer radius of the shell.  Solving for the rotation profile of a shell yields
\begin{equation}
\Omega_s\left(r_0\right)=\left(\frac{3}{2}\frac{\alpha\Delta{E_{orb}}}{M_sr_i^2}\right)^\frac{1}{2}
\end{equation}
where $M_s$ is the shell mass.

We study the limit in which the orbital eccentricity is negligible, resulting in the companion exhibiting Keplerian motion at all radii (\cite{Pollard}).  Fig. \ref{rotfixed} shows envelope rotation profiles for companions with masses ranging from $0.01$ to $0.05$ $M_\odot$.  Also shown is the Keplerian velocity, $v_K$, and the sound speed, $c_s$ inside the envelope.  Higher mass companions supply enough orbital energy to spin-up the envelope above its Keplerian value at a given radius.  The rotational energy will then redistribute via outward mass transfer until Keplerian rotation is re-established.  This effect can be seen in the envelope rotation profile generated by a $0.05$ $M_\odot$ secondary.  It also occurs for larger mass companions, however we do not investigate those as we would expect similar results.  The dashed-dotted vertical line shows the approximate radius at which the companion is tidally shredded and spin-up of the envelope ceases.  Below this depth the companion would form a disk and the mechanism of angular momentum transfer would be different.  In reality the rotation profile should be solved for self-consistently, but this becomes particularly important as soon as the rotational energy exceeds that associated with the sound speed.  We have not incorporated this non-linear effect explicitly, and hence our approach of redistributing the excess rotational energy in the inner regions in Fig. \ref{rotfixed} is crude.

As can be seen from Fig. \ref{rotfixed}, a range of companion masses can produce varying amounts of rotation and differential rotation.  To investigate how this shear energy is deposited, we apply the rotation profiles in Fig. \ref{rotfixed} along with the parameters of the stellar model to our interface dynamo.  For a $0.02$ $M_\odot$ brown dwarf secondary, toroidal field strengths of $\sim1\times10^5$ G are obtained (see Fig. \ref{0.02}).  For $Pr_p=10^{-4}$, the decay of the shear energy and toroidal field are long ($\sim25$ yrs) and occur over several thousand cycle periods.  Therefore, in Fig. \ref{0.02}, the solid line represents the envelope of the dynamo while the smaller insets show a "zoomed-in" region to demonstrate the oscillatory nature of the fields.  For these parameters, the heat generated from turbulent dissipation is greater then the time-integrated Poynting flux, thus we identify this as a thermally induced model.

In Fig. \ref{0.05}, we present a model in which the time-integrated Poynting flux is large enough to unbind the stellar envelope.  In this case, the companion is a $0.05$ $M_\odot$ brown dwarf and spins-up the envelope until it is shredded into a disk.  For this model, $Pr_p=10^{-6}$ with the corresponding Poynting flux decaying in $\sim100$ yrs.  Peak toroidal and poloidal field strengths are comparable to results from Fig. \ref{0.02}, however the lower $Pr_p$ results in less differential rotation energy being converted into heat.  Therefore, we identify this as a magnetically dominated model.

Both magnetically dominated and thermally driven models can be produced for a range of companion masses and diffusion coefficients.  The resultant outflows for the two cases are expected to be qualitatively different.  For interface-dynamo-driven winds, the launching and shaping of the outflow could occur close to the core.  Such an outflow is expected to be collimated, predominately poloidal and aligned with the central rotation axis.  However, if heat is the primary driver mediating the transition from progenitor to PPNe, the resulting outflow is probably quasi-spherical.  Bipolar, magnetically collimated PNe could be the result of our low $Pr_p$, common envelope magnetically dominated models.

\section{Conclusions}
Extraction of rotational energy is likely fundamental to the formation of multipolar PPNe and PNe.  Magnetic dynamos can play an intermediary role in facilitating the extraction of rotational energy.  Here, we have studied a large-scale, dynamical interface dynamo operating in the envelope of a $3.0$ $M_\odot$ AGB star.  The back reaction of field amplification on the shear is included as are the drain of  rotation and differential rotation via both turbulent dissipation (thermally induced) and Poynting flux (magnetically induced).  Two different dynamos are studied: (i) that of an isolated star and (ii) that of a common envelope system in which the secondary is a low-mass companion ($<$ $0.05$ $M_\odot$).

For the isolated case, we find that only when two conditions are met can the single star dynamo drive PPNe.  First, $\Delta\Omega$ must be re-seeded.  This may occur by analogy to the $\lambda$-effect in the sun (\cite{2004muga.book.....R}).  Secondly, $\Omega$ must be constant.  This implies that the field is stored in the shear layer until the end of the AGB phase.  When these two stringent conditions are met, a small fraction of the energy cascade rate can provide the necessary shear energy to sustain the interface dynamo (in some cases as little as $0.1\%$).  Not only is the dynamo maintained, but the time-integrated Poynting flux is large enough to overcome the envelope binding energy.


Whether or not isolated star dynamos can produce a PPNe, a binary interaction can do so more robustly for a wide range of cases.  Common envelope evolution is advantageous in several ways.  Energy and angular momentum are supplied very quickly ($<$ 1 yr) and often in less than or equal to a dynamo cycle period, allowing the dynamo to operate once the secondary has completed its in-spiral.  For our common envelope dynamo model, a range of companion masses can easily supply enough differential rotation energy to power either a dynamo driven jet or a thermally driven outflow.  A magnetically dominated explosion likely produces a collimated, poloidal outflow while that of a thermally induced explosion is expected to be more spherical.  

We have highlighted some of the basic key issues of isolated and common envelope dynamos, however, more detailed research is needed in both areas.  The viability of anisotropic convection reseeding differential rotation must be determined.  If convection cannot reseed shear, then we are faced with the proposition that binary interactions are required to produce axisymmetric PNe.  In the CE case, the complex interaction between companion and envelope, multi-dimensional aspects of the dynamo, realistic rotation profiles of isolated stars and the inclusion of a wider array of secondary masses are just a few of the many problems which warrant future work.  Constraints on the turbulent diffusion coefficient as a function of radius also need to be determined.  The physics involved in transitioning from a dynamo to a fully active jet must be understood.

\acknowledgements{We thank Noam Soker for identifying a needed correction which lead to an  improved manuscript.  We would also like to thank 
Ivan Minchev for useful discussions and comments and Steve Kawaler for use of his evolutionary models.
 JTN acknowledges financial support of a Horton Fellowship from the Laboratory for Laser Energetics through the U. S. Department of Energy and HST grant AR-10972.  
EGB acknowledges support from 
NSF grants AST-0406799, AST-0406823, and NASA grant ATP04-0000-0016
(NNG05GH61G).  AF acknowledges support from (***).

{}
\begin{figure}
\plotone{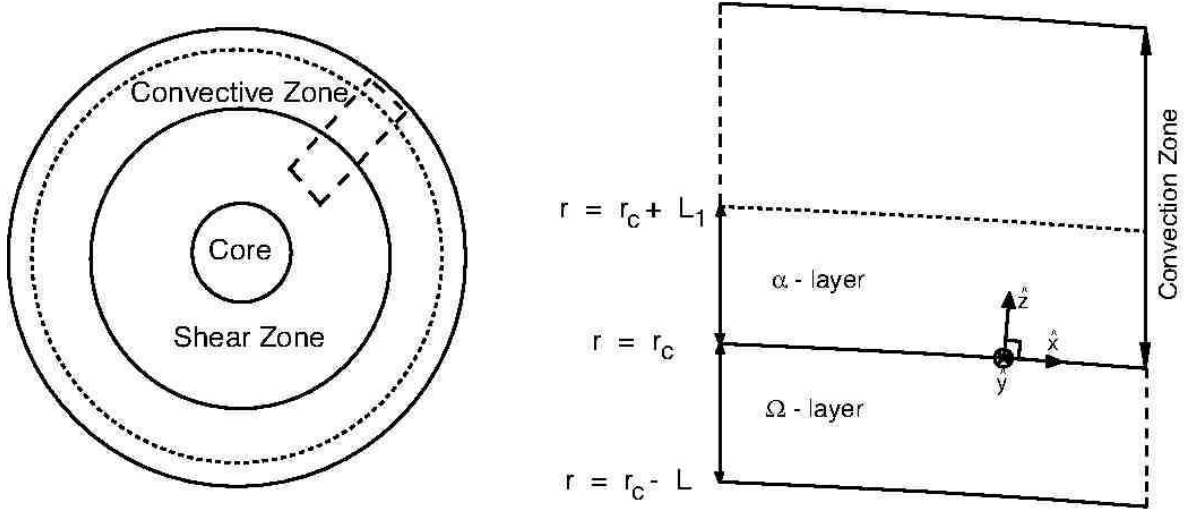}
\caption{A meridonial slice of the dynamo geometry.  The left figure shows the global geometry of the AGB star.  The right figure is a close-up view of the dashed region on the left.  The $\alpha$-effect is driven by convection and occurs in layer of thickness $L_1$ above the differential rotation zone.  The poloidal component of the  field is pumped downwards into the differential zone, where it is wrapped torodially due to the $\Omega$-effect.}
\label{geometry}
\end{figure}

\begin{figure}
\plottwo{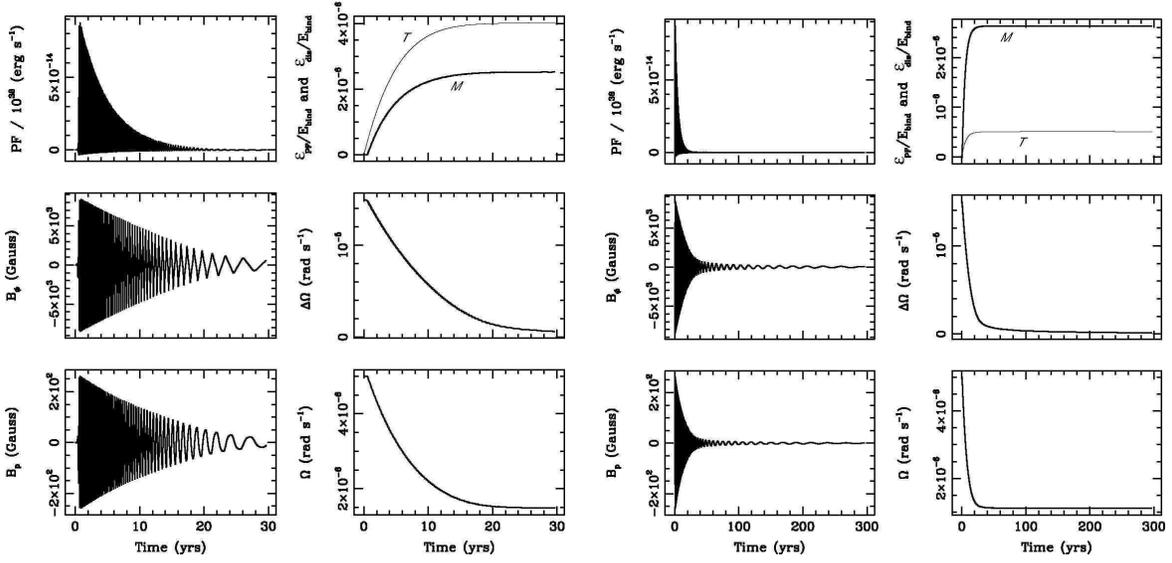}{Nature4.epsf}
\caption{The differential rotation energy is allowed to drain through field amplification and turbulent dissipation.  In this figure, $k_z=5\times10^{-11}$ cm$^{-1}$, $c_p=0.01$, $Q=5.0$.  We define, $\epsilon_{\left[PF,dis\right]}\equiv\int^t_0E_{\left[PF,dis\right]}\left(t'\right)dt'$ and label $M$ ($\epsilon_{PF}$) and $T$ ($\epsilon_{dis}$) on the top right plot to distinguish between the thermal and magnetic contributions to the binding energy.  For the left figure, $c_\phi=10^{-4}$ while the right has $c_\phi=10^{-5}$.  Peak field strengths are a factor of $\sim5-10$ less then those obtained in (\cite{EB2001}).  Differential rotation energy is drained in $<20$ yrs.  Lowering $c_\phi$ results in the differential rotation energy draining at a slower rate, allowing the field to sustain for longer periods of time ($\sim40-50$ yrs).  However, peak field strengths remain the same.}
\label{Nature2}
\end{figure}

\begin{figure}
\plottwo{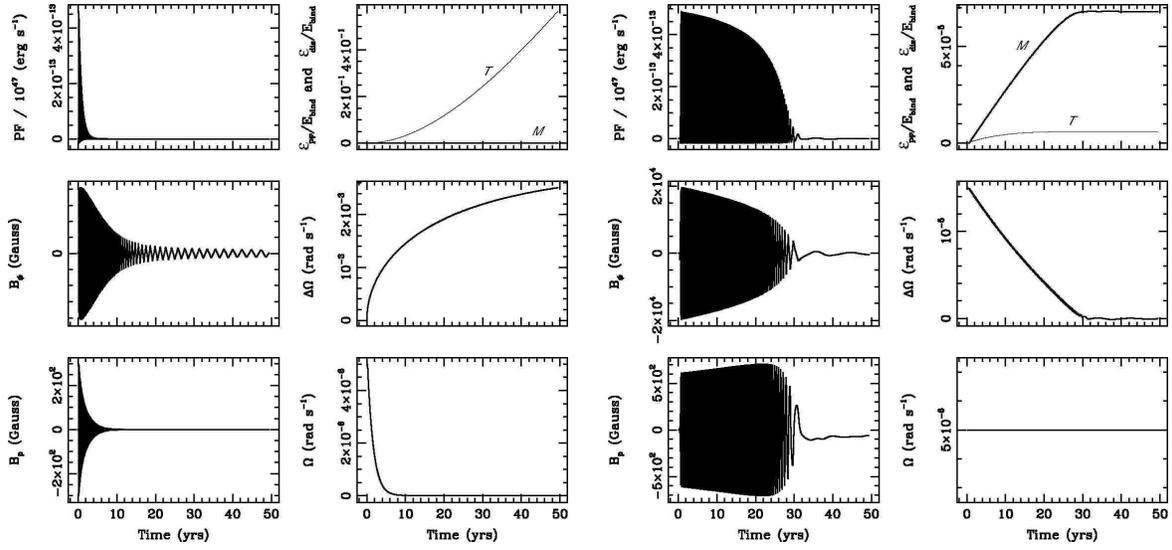}{f_0_omegafixed.epsf}
\caption{Results for reseeding differential rotation through convection.  In the left figure, $f=1$ corresponding to maximum convective resupply.  Rotation is drained through Poynting flux but cannot sustain a dynamically important dynamo.  In the figure on the right, $f=0$ (no resupply of $\Delta\Omega$).  The rotation rate is fixed, corresponding to a buildup of Poynting flux in the interface layer.}
\label{reseed1}
\end{figure}

\begin{figure}
\plotone{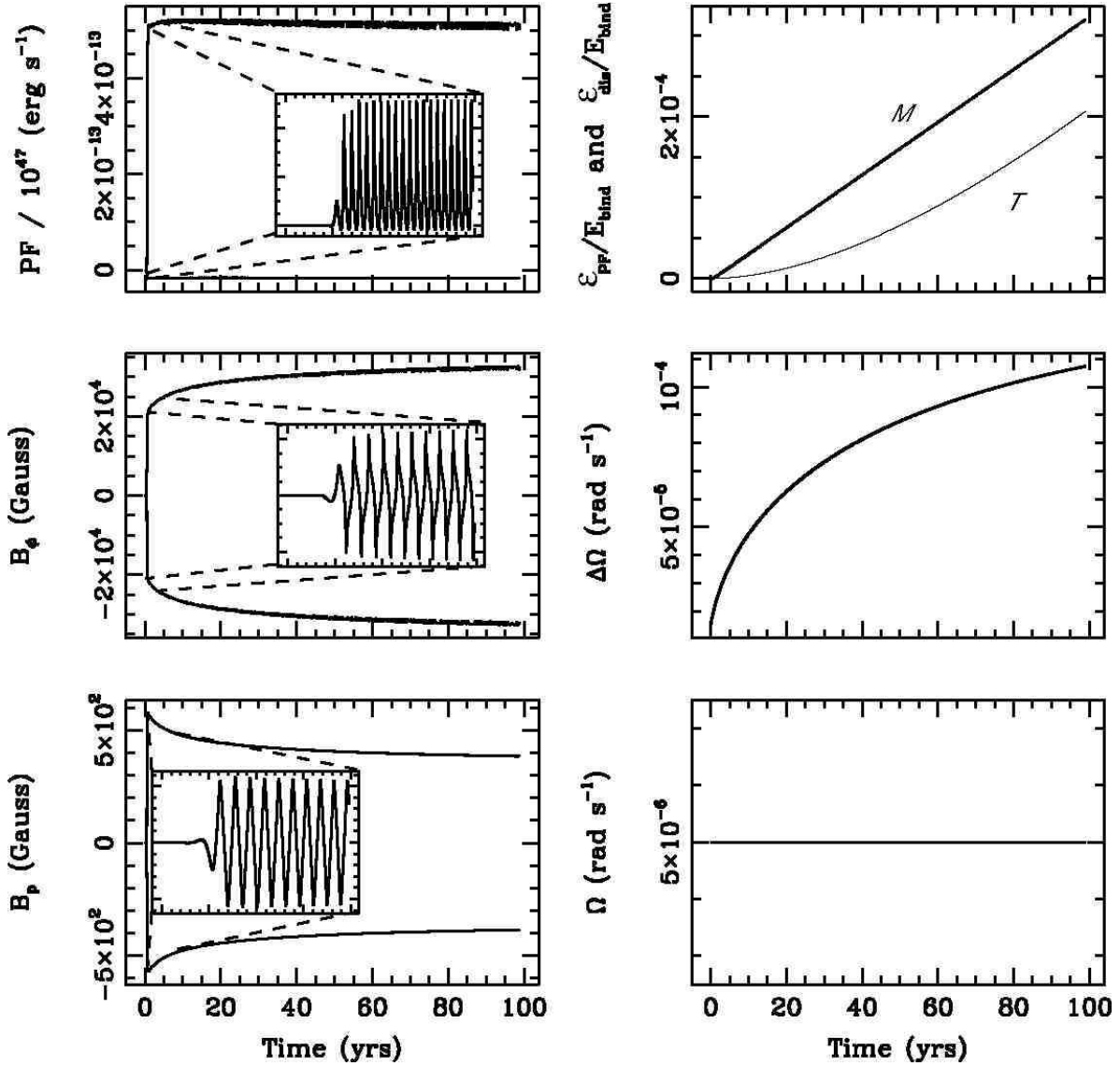}
\caption{Convective resupply results in a steady-state differential rotation profile.  For the left column, the envelope of the poloidal, toroidal and Poynting flux is plotted.  The Poynting flux is sustained at $\sim5\times10^{34}$ erg/s.  The sustained Poynting flux supplies enough energy to unbind the envelope of our $3$ $M_\odot$ model at the end of the AGB phase ($\sim10^5$ yrs).  In this figure, $c_\phi=10^{-5}$ and $f=10^{-3}$ implying that only $\sim0.1\%$ of the cascade energy must be converted into differential rotation energy to supply the requisite Poynting flux.  This model predicts a magnetically dominated explosion.}
\label{reseed2}
\end{figure}

\begin{figure}
\plottwo{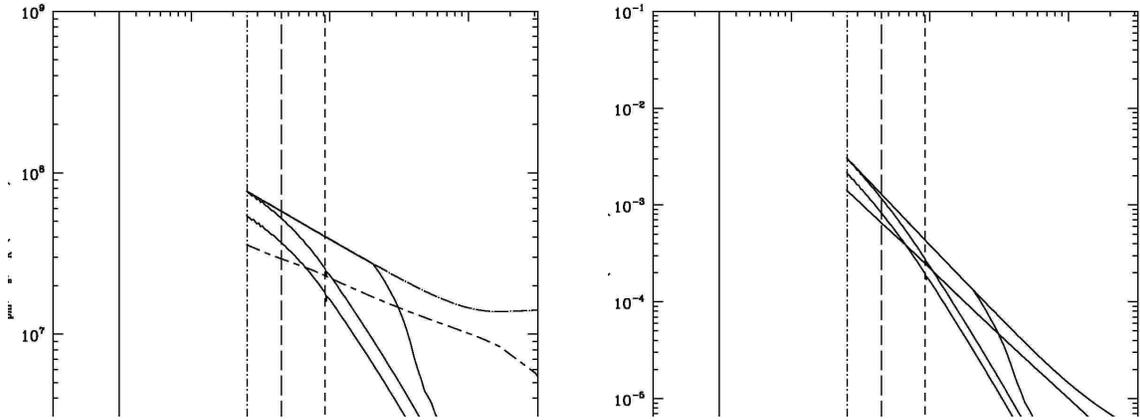}{omegaafter.epsf}
\caption{Rotation profiles generated from the transfer of angular momentum from companion to envelope in our AGB star.  In both figures, the solid curves represent the resulting profiles for companions of masses 0.05 (top), 0.02 and 0.01 (bottom) $M_\odot$.  For the left figure, the dashed curve represents the Keplerian velocity while the dashed-dotted curve is the sound speed.  The 0.05 $M_\odot$ companion initially spins up the envelope such that the inner region is rotating faster then the Keplerian velocity.  Mass redistribution ensues and transfers matter outward until the rotation profile drops below Keplerian.  The right figure presents the angular velocity corresponding to the left figure.  The dash-dot vertical line is the approximate radius at which the companion is tidally shredded.  The large-dash vertical line is the boundary of the shear layer in \cite{EB2001} while the small dash line is the base of the convection zone.  These profiles assume that $\alpha=0.3$.}
\label{rotfixed}
\end{figure}

\begin{figure}
\plotone{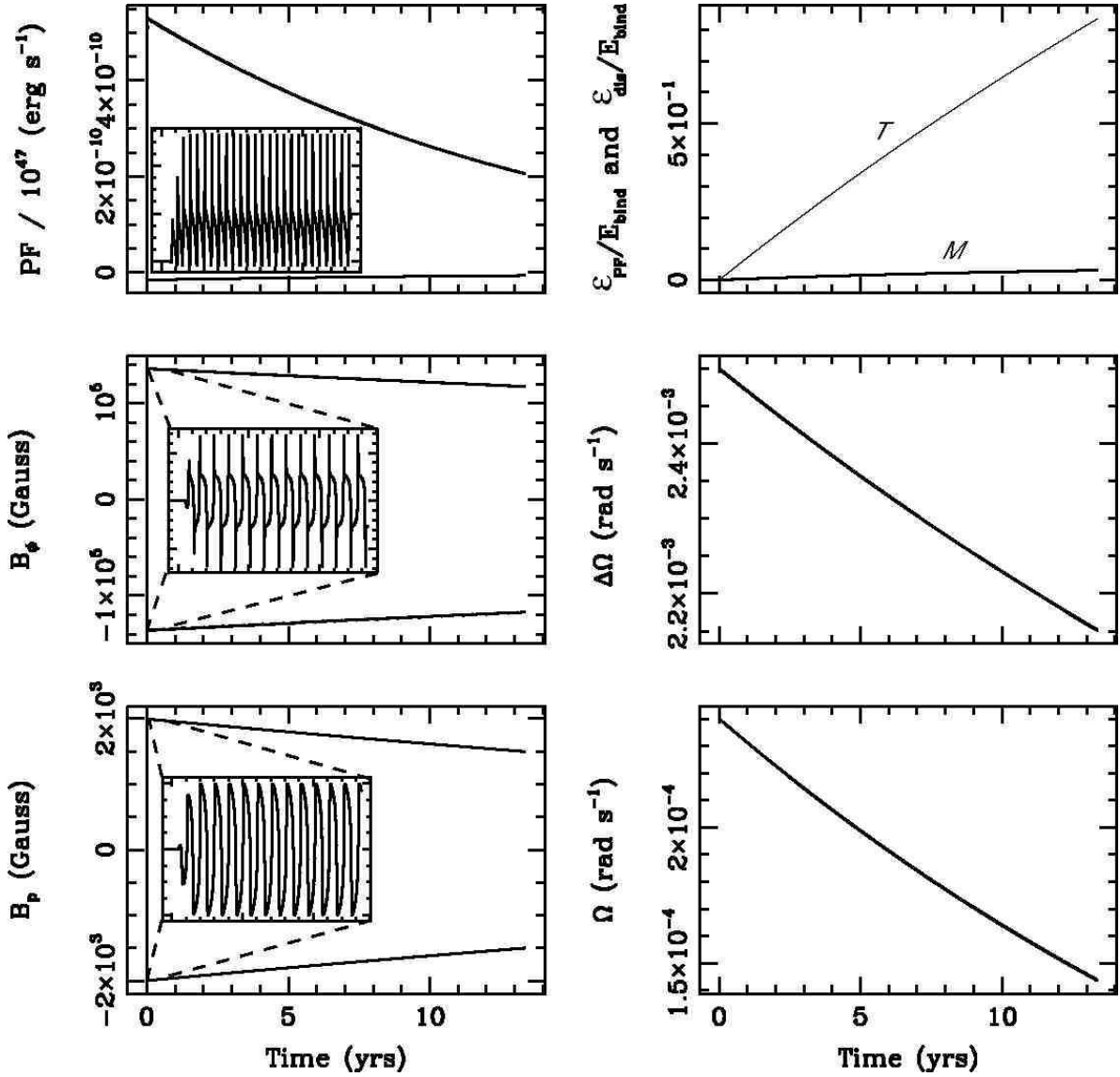}
\caption{Interface dynamo resulting from the in-spiral of a $0.02$ $M_\odot$ brown dwarf in the interior of our model AGB star.  The differential rotation zone extends from the base of the convection zone to the radius at which the secondary is tidally shredded (\cite{JN2006}).  In this model, $P_M=10^{-4}$ and $Q=5$, $\Omega_0=2.3\times10^{-4}$ rad/s, $\Delta\Omega_0=2.5\times10^{-3}$ rad/s and $\delta/L=1$.  In the left column, the envelope of the Poynting flux (top), toroidal field (middle) and poloidal field (bottom) are drawn with a solid line.  The insets represent the time evolution from 0 to 0.2 yrs.  The 
vertical scale of the insets are the same as the corresponding larger figure.}
\label{0.02}
\end{figure}

\begin{figure}
\plotone{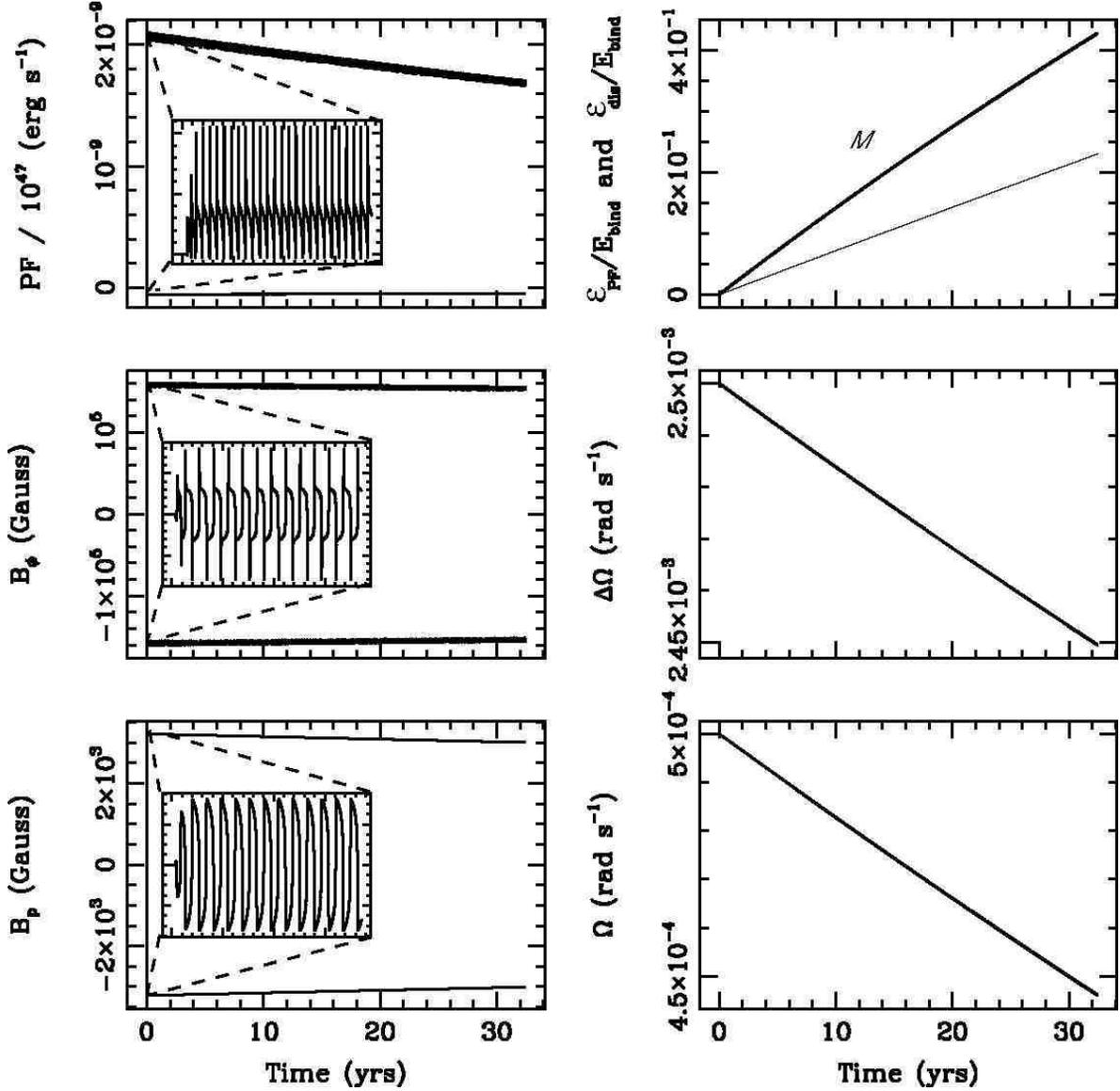}
\caption{Interface dynamo resulting from the in-spiral of a $0.05$ $M_\odot$ brown dwarf.  In this model $P_M=10^{-6}$, $Q=5$, $\delta/L=1$, $\Omega_0=5\times10^{-4}$ rad/s and $\Delta\Omega_0=2.5\times10^{-3}$ rad/s.  The insets represent the time evolution from 0 to 0.2 years.  The vertical scale of the insets are the same as the corresponding larger figure.}
\label{0.05}
\end{figure}

}


\clearpage
\begin{thebibliography}{}

\bibitem[Bains et al. (2004)]{2004MNRAS.354..529B} Bains, I., Richards, 
A.~M.~S., Gledhill, T.~M., \& Yates, J.~A.\ 2004, \mnras, 354, 529 

\bibitem[Balick \& Frank (2002)]{2002ARA&A..40..439B} Balick, B., \& Frank, 
A.\ 2002, \araa, 40, 439 

\bibitem[Blackman et al. (2001)]{EB2001} 
Blackman, E. G., Frank, A., Markiel, J. A., Thomas, J. H. \& Van Horn, H. M. 2001 Nature, 409, 485

\bibitem[Blackman \& Field (2002)]{2002PhRvL..89z5007B} Blackman, E.~G., \& 
Field, G.~B.\ 2002, Physical Review Letters, 89, 265007 

\bibitem[Blackman (2004)]{2004ASPC..313..401B} Blackman, E.~G.\ 2004, in ASP 
Conf.~Ser.~313: Asymmetrical Planetary Nebulae III: Winds, Structure and 
the Thunderbird, 313, 401 Edited by Margaret Meixner, Joel H. Kastner, Bruce Balick and Noam Soker.

\bibitem[Blackman, Nordhaus \& Thomas (2006)]{BNT}
Blackman, E. G., Nordhaus, J. T. \& Thomas, J. H. 2006 New Astronomy, 11, 452

\bibitem[Brandenburg \& Subramanian (2005)]{2005PhR...417....1B} 
Brandenburg, A., \& Subramanian, K.\ 2005, \physrep, 417, 1 

\bibitem[Browning et al. (2006)]{2006ApJ...648L.157B} Browning, M.~K., 
Miesch, M.~S., Brun, A.~S., \& Toomre, J.\ 2006, \apjl, 648, L157

\bibitem[Burleigh et al. (2006)]{2006astro.ph..9366B} Burleigh, M.~R., 
Hogan, E., Dobbie, P.~D., Napiwotzki, R., \& Maxted, P.~F.~L.\ 2006, ArXiv 
Astrophysics e-prints, arXiv:astro-ph/0609366

\bibitem[De Marco et al. (2004)]{2004ApJ...602L..93D} 
De Marco, O., Bond, H.~E., Harmer, D., \& Fleming, A.~J.\ 2004, \apjl, 602, L93 

\bibitem[Dikpati et al. (2006)]{2006ApJ...638..564D} Dikpati, M., Gilman, 
P.~A., \& MacGregor, K.~B.\ 2006, \apj, 638, 564

\bibitem[Etoka \& Diamond (2004)]{2004MNRAS.348...34E} Etoka, S., \& 
Diamond, P.\ 2004, \mnras, 348, 34 

\bibitem[Iben \& Livio (1993)]{1993PASP..105.1373I} 
Iben, I.~J., \& Livio, M.\ 1993, \pasp, 105, 1373

\bibitem[Jordan et al. (2005)]{2005A&A...432..273J} 
Jordan, S., Werner, K., \& O'Toole, S.~J.\ 2005, \aap, 432, 273

\bibitem[Kleeorin \& Ruzmaikin (1982)]{Kleeorin1982}
Kleeorin, N., \& Ruzmaikin, A. A. 1982, Magnetohydrodynamics 18, 116

\bibitem[Markiel et al. (1994)]{1994ApJ...430..834M} Markiel, J.~A., Thomas, 
J.~H., \& van Horn, H.~M.\ 1994, \apj, 430, 834 

\bibitem[Matt et al. (2004)]{2004ASPC..313..449M} Matt, S., Frank, A., \& 
Blackman, E.~G.\ 2004, ASP Conf.~Ser.~313: Asymmetrical Planetary Nebulae 
III: Winds, Structure and the Thunderbird, 313, 449 Edited by Margaret Meixner, Joel H. Kastner, Bruce Balick and Noam Soker.

\bibitem[Mauron \& Huggins (2006)]{2006A&A...452..257M} Mauron, N., \& 
Huggins, P.~J.\ 2006, \aap, 452, 257 

\bibitem[Maxted et al. (2006)]{2006Natur.442..543M} Maxted, P.~F.~L., 
Napiwotzki, R., Dobbie, P.~D., \& Burleigh, M.~R.\ 2006, \nat, 442, 543

\bibitem[Moe \& De Marco (2006)]{2006astro.ph..6354M} Moe, M., \& De Marco, 
O.\ 2006, ArXiv Astrophysics e-prints, arXiv:astro-ph/0606354 

\bibitem[Moffatt (1978)]{1978mfge.book.....M} Moffatt, H.~K.\ 1978, 
Cambridge, England, Cambridge University Press, 1978.~353 p.,

\bibitem[Nordhaus \& Blackman (2006)]{JN2006} Nordhaus, J., \& 
Blackman, E.~G.\ 2006, \mnras, 370, 2004 

\bibitem[Paczynski (1976)]{1976IAUS...73...75P} 
Paczynski, B.\ 1976, IAU Symp.~ 73: Structure and Evolution of Close Binary Systems, 73, 75


\bibitem[Parker (1955)]{1955ApJ...121..491P} Parker, E.~N.\ 1955, \apj, 121, 
491 


\bibitem[Parker (1979)]{1979cmft.book.....P} Parker, E.~N.\ 1979, Oxford, 
Clarendon Press; New York, Oxford University Press, 1979, 858 p.,  

\bibitem[Parker (1993)]{1993ApJ...408..707P} Parker, E.~N.\ 1993, \apj, 408, 
707 

\bibitem[Pascoli (1993)]{1993JApA...14...65P} Pascoli, G.\ 1993, Journal of 
Astrophysics and Astronomy, 14, 65 

\bibitem[Pascoli (1997)]{1997ApJ...489..946P} Pascoli, G.\ 1997, \apj, 489, 
946 

\bibitem[Pollard (1979)]{Pollard}
Pollard, H. 1979, \textit{Celestial Mechanics} (Mathematical Association of America)


\bibitem[Robinson \& Durney (1982)]{1982A&A...108..322R} Robinson, R.~D., \& 
Durney, B.~R.\ 1982, \aap, 108, 322

\bibitem[R{\"u}diger (1989)]{1989QB523.R93......} Rudiger, G.\ 1989, New York : 
Gordon and Breach Science Publishers, c1989.

\bibitem[R{\"u}diger \& Hollerbach (2004)]{2004muga.book.....R} R{\"u}diger, 
G., \& Hollerbach, R.\ 2004, The Magnetic Universe: Geophysical and 
Astrophysical Dynamo Theory, by G{\"u}nther R{\"u}diger, Rainer Hollerbach, 
pp.~343.~ISBN 3-527-40409-0.~Wiley-VCH , August 2004.

\bibitem[Shu (1992)]{1992phas.book.....S} Shu, F.~H.\ 1992, Physics of 
Astrophysics, Vol.~II, by Frank H.~Shu.~Published by University Science 
Books, ISBN 0-935702-65-2, 476pp, 1992.

\bibitem[Soker \& Zoabi (2002)]{2002MNRAS.329..204S} Soker, N., \& Zoabi, 
E.\ 2002, \mnras, 329, 204

\bibitem[Soker (2006)]{2006PASP..118..260S} 
Soker, N.\ 2006, \pasp, 118, 260

\bibitem[Sorensen \& Pollacco (2004)]{2004ASPC..313..515S} 
Sorensen, P., \& Pollacco, D.\ 2004, ASP Conf.~Ser.~313: Asymmetrical Planetary Nebulae III: 
Winds, Structure and the Thunderbird, 313, 515 Edited by Margaret Meixner, Joel H. Kastner, Bruce Balick and Noam Soker.

\bibitem[Thomas et al. (1995)]{1995ApJ...453..403T} Thomas, J.~H., Markiel, 
J.~A., \& van Horn, H.~M.\ 1995, \apj, 453, 403

\bibitem[Tout \& Pringle (1992)]{1992MNRAS.256..269T} Tout, C.~A., \& 
Pringle, J.~E.\ 1992, \mnras, 256, 269 

\bibitem[Vlemmings et al. (2006)]{2006Natur.440...58V} Vlemmings, W.~H.~T., 
Diamond, P.~J., \& Imai, H.\ 2006, \nat, 440, 58 












 





 







 








 

\end{thebibliography}
\end{document}